# Pseudoepitaxial transrotational structures in 14 nm-thick NiSi layers on [001] silicon

**Alessandra Alberti[1], Corrado Bongiorno[1], Brunella Cafra[1], Giovanni Mannino[1], Emanuele Rimini[1], Till Metzger[2], Cristian Mocuta[2,] Thorsten Kammler[3], Thomas Feudel[3]**

[1]CNR-IMM, Sezione di Catania, Stradale Primosole50, 95121 Catania, Italy

[2]European Synchrotron Radiation Facility, BP 220, 38043 Grenoble Cedex, France

[3]AMD Saxony LLC & Co. KG, Wilschdorfer Landstrasse 101, Dresden,Germany

**Abstract**

In a system consisting of two different lattices, the structural stability is ensured when an epitaxial relationship occurs between them and allows the system to retain the stress, avoiding the formation of a polycristalline film. The phenomenon occurs if the film thickness does not exceed a critical value. Here we show that, in spite of its orthombic structure, a 14nm-thick NiSi layer can three-dimensionally (3D) adapt to the cubic Si lattice by forming *transrotational* domains. Each domain arises by the continuous bending of the NiSi lattice, maintaining a close relationship with the substrate structure. The presence of *transrotational* domains does not cause a roughening of the layer but instead it improves the structural and electrical stability of the silicide in comparison with a 24nm-thick layer formed using the same annealing process. These results have relevant implications on thickness scaling of NiSi layers currently used as metallizations of electronic devices.

**Correspondence should be addressed to :** Alessandra Alberti ( Alessandra.Alberti@imm.cnr.it)

**INTRODUCTION**

Low resistivity Nickel silicide (NiSi) will replace cobalt silicide in next generation metal-oxide-semiconductor devices (MOSFET). The main advantage of NiSi is the reduction of Si consumption without an increase of the sheet resistance in order to ensure shallow junction integrity and low contact resistance. In this respect, it is mandatory to avoid the transition of NiSi to the high resistivity $NiSi_2$ phase as the annealing temperature exceeds 700-800°C [1,2,3] and to control the interfacial properties [4,5,6,7,8]. A general method to optimise the structure of the silicide layer consists of reducing the interfacial free energy between the layer and the substrate and/or increasing the volume energy gain of the nucleation barrier [9]. Recently, it has been shown that, in spite of the lattice dissimilarity of the orthorombic NiSi with cubic Si, an ordered relationship between the silicide layer and the substrate is established; this peculiar growth was called axiotaxy [10,11]. Axiotaxy represents an intermediate case between heteroepitaxy and a random growth, it is a fibre-like texture but with an off-normal fibre axis resulting in a one-dimensional (1D) periodic interface.

In this work we show that the NiSi lattice can three-dimensionally (3D) adapt to the Si lattice by forming *transrotational* domains. This phenomenon has been observed for reaction temperatures between 260 and 900°C in pure nitrogen and in vacuum, by using rapid thermal annealing (spike or 30 sec) and furnace annealing, on Cz-Si and Silicon On Insulator (SOI) substrates. The resulting structural properties of the silicide layer are promising to improve the structural and electrical stability of NiSi.

**EXPERIMENTAL**

On [001] Cz-Si cleaned substrates, 7 nm-thick Nickel layers were deposited by sputtering, and subsequently annealed in a Rapid Thermal Annealer (RTA) for 30s or by spike annealing in pure nitrogen ambient between 450 and 900°C, or by furnace annealing in the range between 260 and 350°C. The reacted samples were analysed by X-ray diffraction (XRD), Transmission Electron Microscopy (TEM) and Selected Area Diffraction (SAD) analyses. Pole figures were measured at the ID01 beamline of the European Synchrotron Radiation Facilities (ESRF) in Grenoble using photon energy of E=8 keV (KαCu) to study the three-dimensional configuration of the NiSi lattice with respect to the substrate.

**RESULTS AND DISCUSSION**

Here we report on the structure of a 14-nm NiSi layer formed after 550°C spike annealing as representative of all the samples described above  Figure 1a is a large area plan-view TEM image of the silicide layer taken at 5° tilt with respect to the [001] Si direction to reduce the substrate contributions. The pattern does not match at all with that of a conventional polycrystalline film since the boundaries between adjacent domains are not clearly defined. On the contrary, an intricate network of extinction contours covers the entire area of the sample. The extinction contours create some peculiar, highly ordered and symmetric structures defined by the intersection of more than two fringes, as indicated by the dashed ellipses in figure 1a. A thicker silicide layer (24 nm), formed under the same annealing conditions, has, instead, the structure of a conventional polycrystal, as shown in figure 1c and d.

A network of extinction contours was first observed by Kolosov[12] in crystalline Se or Fe-based spherulites in an amorphous matrix, and subsequently in the crystallization of amorphous chalcogenides films[13,14,15]. In those works, it is assessed that the domains, defined by the intersection of extinction contours, were generated by a continuous rotation of the crystal that bends without producing roughening of the interfaces (*transrotational structures*)[26]. Likewise, we have found that within each domain the lattice of NiSi bends around a fixed zone axis following an almost hemispherical path. The centre of the domain, at the intersection of the extinction contours, is the only region in which the lattice is in the axis condition. It has been observed that, by tilting the sample, the centre of the intersection between the contours moves in the same direction as the tilt, and from this property it is possible to determine the sign of the curvature of these domains. It has been found that the domains are concave. As a difference with respect to the chalcogenides[29], our films have a tight relationship with the substrate which is not amorphous but crystalline.

The electron diffraction pattern shown in figure 1b is characterised by two intense spots, perpendicular to each other, aligned to the [220] directions of the substrate. They are due to the unique (020) planes of the NiSi orthorombic lattice and therefore they belong to different zone axes rotated by 90°.

To fully characterise the relationship between the film and the substrate, SAD analyses have been performed on different domains of the sample. It is noteworthy that all the collected diffraction pattern can be associated only to three different zone axes, i.e. the [101], [102] and [201] directions of the NiSi lattice. Each of those zone axes has been correlated to a particular domain, characterised by different number and mutual position of the extinction contours. The last two zone axes have similar diffraction patterns and therefore the associated domains are not easily distinguishable in the plan-view image. The main domains, identified as type-I (Fig. 2a) and type-II (Fig. 2b), are formed by the intersection of four or three fringes respectively (type-III not shown). It is worth to note that the extinction contours are almost perpendicular to the diffraction spots labelled in the corresponding SAD (right side of fig.2), and this confirms the *transrotational* structure of the domains. All the intense spots in figure 1b are reproduced by these patterns and by those rotated by 90°. This implies that, within the large area analysed, the film mainly shows only three types of domains orientations. The peculiarity of this kind of growth is the close relationship between the lattices of NiSi and Si in which, within each bending contour of the domain, the (020) planes of the silicide are anchored at the interface with silicon to its (220) planes. A schematic view of the planes bending along one direction is sketched in figure 2c.

A ±18° continuous rotation of the orthorombic NiSi lattice around the direction normal to the (020) planes moves the lattice configuration from the [102] to the [201] zone axis passing through [101] without meeting other relevant axes in between. This result, combined with the presence of the bending contours within the domains, suggests that along the (020) related fringes the lattice of NiSi is able to bend in a continuous way preserving the (020) planes aligned to the (220) planes of the substrate. This kind of preferential alignment is lost immediately outside the (020) fringes in all the other directions, as sketched in figure 2c. Therefore, the continuous rotational configurations of the NiSi lattice around the (020) planes, from one zone axis to another, are all present within a single *transrotational* domain instead of being associated to a collection of differently oriented domains

In order to investigate the spatial distribution of these preferentially oriented fringes, dark-field TEM analyses were done over the sample area shown in figure 1a, by selecting each of the two [020] spots of the diffraction pattern labelled in figure 1b. Two clearly distinguishable and spatially separated set of bright fringes are detected over the entire sample area. The first set of fringes, shown in Fig.3a, covers the upper part of the sample area; the second set of fringes, rotated by 90°, occupies the lower part of the analysed area (Fig.3b). These planes follow the symmetry of (220) planes of silicon.

The statistical angular distribution of the (020) NiSi planes was obtained by X-ray diffraction analysis changing the azimuthal angle $\phi$ from 0 to 360°, and the polar angle $\chi$ from 0 to 86°. The resulting stereo projection is shown on a logarithmic colour scale in figure 4a. The intensity distribution consists of well defined regions placed in symmetrical positions. For the selected wavelength and Bragg angle, the pole figure contains contributions from the (020) and (013) planes of the NiSi film and the (311) planes of the substrate, having all similar d-spacing (d=1.629, 1.632, and 1.637Å respectively). The signals of the substrate are identified by the dark squares (fig.4a) and are located at $\chi=26°$ and $\chi=72°$ with a distribution along $\phi$ which follows the symmetry of the silicon lattice. The remaining intense patterns are due to the NiSi layer. Their distribution and shape cannot be accounted for either a random growth or epitaxy or axiotaxy. Random growth should, in fact, result in a uniform angular intensity distribution; epitaxy should produce a small rounded pattern in correspondence of substrate planes; axiotaxy should give circular features in the spherical representation around $\chi=45°$ and $\phi=45, 135, 225$ and $315°$ [24,25]. It is noteworthy, instead, the presence of narrow elongated patterns at ϕ=0, 90,180, 270° that follow the symmetry of (220) Si planes. They extend by only 4° along ϕ and from 86 to 70° along χ. These contributions have been related to the (020) planes of NiSi. Other extended features are found in proximity of χ=45° and ϕ=45, 135, 225 and 315°. Each of them has the shape of a lozenge, as shown in quadrant IV of the pole figure, and is related to (013) planes of NiSi. The sketch in figure 4b, representing a row of domains, can be used to explain the distribution and shape of all the diffraction patterns attributed to the NiSi lattice. Starting by the core of the type-I domain, labelled with "A" and corresponding to the [101] zone axis, the core "C" of a type-II domain is obtained by a continuous rotation of the NiSi lattice around the direction normal to the (020) planes. These planes indeed remain in the same position. The rotation causes, instead, all the (013) planes to move according to their mutual position in the NiSi lattice and therefore to contribute to the diffraction pattern in different positions. The related diffracted spots are represented in quadrants I and II of the pole figure, and they move from the black to the blue positions passing through the red positions. These movements of the lattice do not produce any shift of the (020) spots which remain at ϕ=0° and χ=90° (black, red and blue circles are superimposed). Following the *transrotational* structure of each domain (fig.2), others representative lattice configurations are obtained and represented with the corresponding colours. When the lattice lies in the "E", "F", "D", "G" configurations, the (020) spot moves along χ (yellow circle) or slightly apart from this axis (the other colours). Some representative configurations within the type-III domains have also been considered ("C', D', F' ") and represented as dashed circles in quadrant. II According to the results shown in figure 3, the contributions of the same row of domains rotated by 90° have additionally been represented in quadrant III. In this way, half of the lozenge-shaped pattern is covered by the coloured circles, the other symmetric side being completed by representing the right part of the domains (not superimposed). The corresponding (020) spots tend to cover the elongated patterns at the edge of the pole figure describing, indeed, the bending of the (020) planes outside the related extinction contour. The higher intensity of the central part of these patterns is due to the size and shape of the domains. The analysis of this pole figure proves that the location of the (013) planes of NiSi are in tight relationship with the *transrotational* structures of the domains and also that these planes assume a continuous but limited number of configurations.

To further support and extend these results, the pole figure of the (202)/(211) planes of NiSi has been analysed with the appropriate choice of the Bragg angle, and the results are shown in figure 5. These two set of planes, and also the (220) planes of silicon, are undistinguishable due to their similar d-spacing. This pole figure is characterised by four extended patterns located at $\phi=0$, 180, 270 and 360°, which have been attributed to (202) planes of NiSi. Other four symmetrical lozenge-shaped contributions, centred at $\chi=45°$ and $\phi=45$, 135, 225 and 315°, have been found, each having two splitting arms which extend towards the edge of the figure. The core of the these patterns is due to (220) Si planes, as indicated by the black squares in figure 5. The further extension of these large patterns unambiguously establishes the presence of continuous contributions from the (211) planes of the silicide layer. Following the domain structure sketched on the left side of figure 5, some representative contributions given by (211) and (202) NiSi planes are represented in the pole figure by coloured circles. They also represent contributions from a similar row of domains rotated by 90° (see figure 3). The diffracted spots accumulate within the lozenge-shape features and also reproduce their splitting arms. In correspondence, the (202) planes move along and in proximity of the $\chi$ axis as a consequence of the (020) planes bending outside the related bending contours.

Our findings significantly support the *transrotational* nature of the domains, and moreover they assess that the entire NiSi layer has grown following these structures. In addition, the analysis of the poles introduces a new description of the growth of NiSi which substantially differs from axiotaxy [24]. Let's consider the three-dimensional pseudomatch sketched on the right side of figure 5, in which a double relationship between the lattices of NiSi and Si is established. It consists of a couple of (211) planes of NiSi which lie, except for a tilt in $\chi$, on the (220) planes of Si which cross the diagonal of the cube faces. The positions of these NiSi planes in the stereo projection correspond to the blue circles of figure 5, where the diffracted intensity is high. Therefore, this very peculiar configuration is real and extremely frequent.

**CONCLUSIONS**

It has been shown that NiSi tightly adapt to silicon by means of *transrotational* domains. Within the domains, the NiSi lattice establishes a double ordered relationship with silicon, in which (211) NiSi planes tend to match (220) Si planes in proximity of $\chi=45°$ and $\phi=45, 135, 225$ and $315°$. But, due to the not allowed match between the orthorhombic and the cubic lattice, the NiSi lattice has a certain degree of "distortion" with respect to Si along $\chi$ and $\phi$. The measure of these distortions is given by the bending of the domain and, therefore, by the extension of the related diffraction patterns in the pole figures. Each NiSi domain has, indeed, grown following the trace of the substrate, and this phenomenon is related to the fact that Ni is the diffusing species into Si to form NiSi.

In the light of our finding, we believe that a driving force to grow these *transrotational* domains is the gain in the volume term rather than in the surface energy contribution. The Ni thickness becomes a critical parameter to tailor the growth process of the silicide since we do not observe *transrotational* domains for a 24 nm-thick layer. In this respect, the kinetics of NiSi formation vs. layer thickness is still under investigation since the formation of transrotational domains prospects relevant implications on the scaling of NiSi layers currently used as metallizations of electronic devices.

**Acknowledgements:** This work has been partially supported by the European Project Impulse (IST-2001-32061)

**FIGURE CAPTIONS**

**Figure 1**    **Structural analyses.** a) Plan-view TEM image of a 14nm-thick silicide layer reacted by spike annealing at 550°C. An intricate network of extinction contours (black lines) covers the entire area of the sample. Domains are defined by the intersection of more than two of those contours. b) Large area diffraction analysis showing the location of (020) planes of NiSi with respect to the Si lattice structure. These planes, which are perpendicular to the sample surface, have only two allowed configurations at 90° one to each other (see the dashed circles). C) and d) : plan-view image and large area diffraction analysis of a 24nm-thick NiSi layer obtained by spike annealing at 550°C..

**Figure 2**    **Domain structure.** Details of type–I (a) and type–II (b) domains defined on the basis of the corresponding SAD patterns. The extinction contours are almost perpendicular to the diffraction spots labelled in the corresponding SAD, and in these regions the NiSi planes are in Bragg conditions. This conditions is lost outside the contours due to the continuous bending of the NiSi planes as sketched in (c). The bending occurs in all the directions, as represented in the case of (020) and (202) NiSi planes in (d), and it changes the properties of the interface without producing structural roughnening of the layer.

**Figure 3**    **(020) NiSi bending contours.** Dark-field TEM analyses obtained by selecting each of the outlined [020] spots in figure 1b. The upper part of the sample has domains with similarly oriented (020) bending contours (a). The domains in the lower part have the (020) bending contours rotated by 90° (b). Therefore, two regions are defined on the basis of the orientation of these contours, and they cover the entire area of the sample.

**Figure 4**    (a) (020)/(013) NiSi pole figure. Contributions from (131) planes of Si are also present (black squares). The lozenge-shaped patterns at $\chi=45°$ are due to (013) NiSi planes and the elongated patterns at the edges of the figure are due to the (020) planes of NiSi. (b) Schematic view of the domains forming the NiSi layer, with the corresponding diffraction patterns to identify the planes orientation. In the core of the type-I domain ("A"), the (020) planes contribute to the pole figure at χ=90° φ=0°, while the diffracted spots from (013) planes lie within the lozenges of -quadrants I and II (black circles). Moving from "A" to another regions of the domains, the diffracted spots distribute as shown by the corresponding coloured circles. Note that they accumulate in the region of the lozenges and at the edges of the figure according to the symmetry of the substrate.

**Figure 5** (a) (202)/(211) NiSi pole figure. Contributions of Si (220) planes are also present (black squares). Following the domain structure (left side schematic), some representative contributions from the NiSi lattice have been represented in the pole figure as coloured circles. Note that the (202) planes of NiSi tend to cover the features at the edge of the figure and the contribution from (211) planes accumulated within the lozenge in proximity of the Si signals and also along the splitting arms. Due to the mutual position of these planes within the lattice of NiSi, the relationship with Si is such that when one of these NiSi planes approaches the core of the lozenge, the other in the near quadrant moves quite far away. This extreme configuration is not convenient. Instead, one of the most convenient location of NiSi planes with respect to the Si lattice is sketched on the right side of the figure.

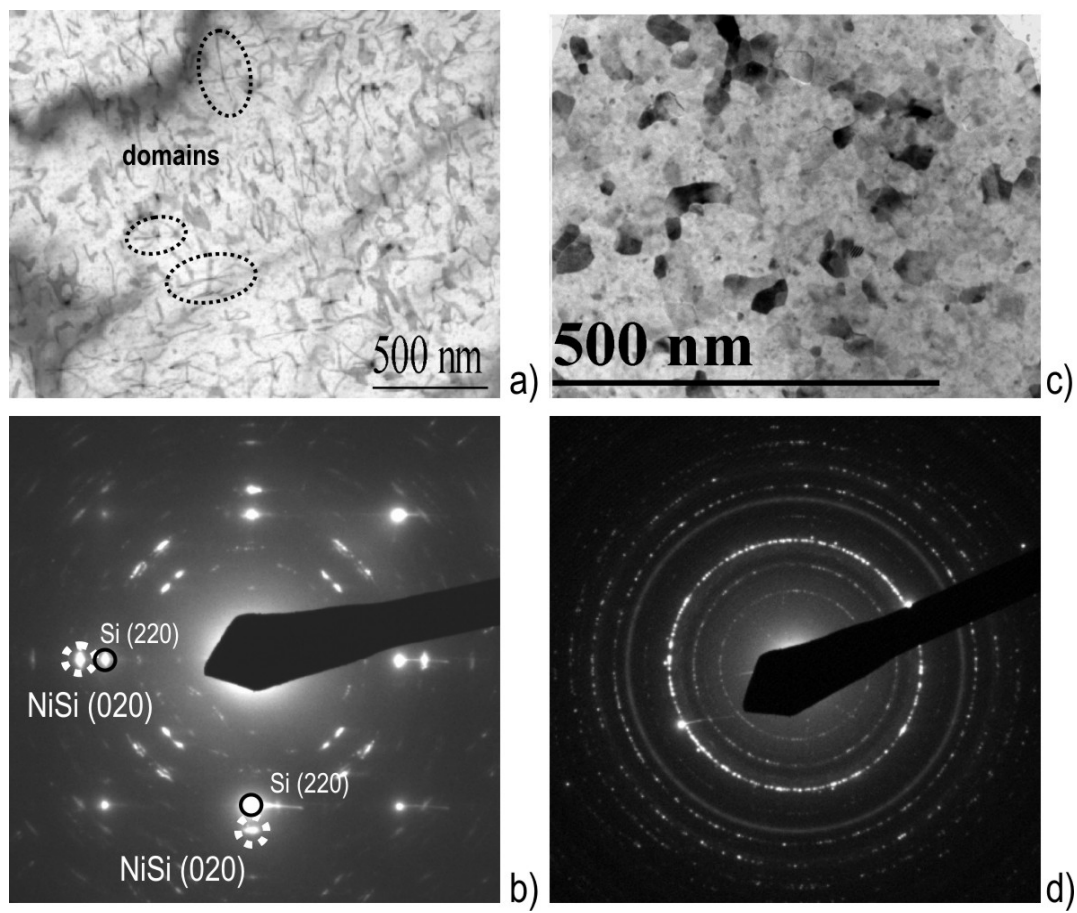

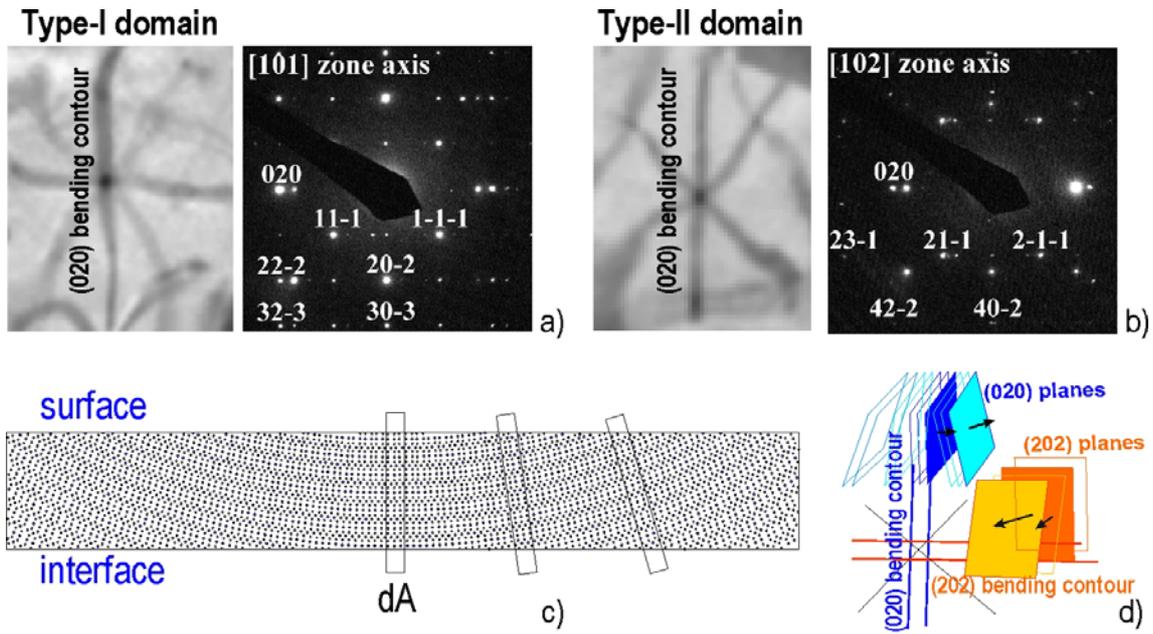

A. Alberti et al.
FIGURE 2 OF 5

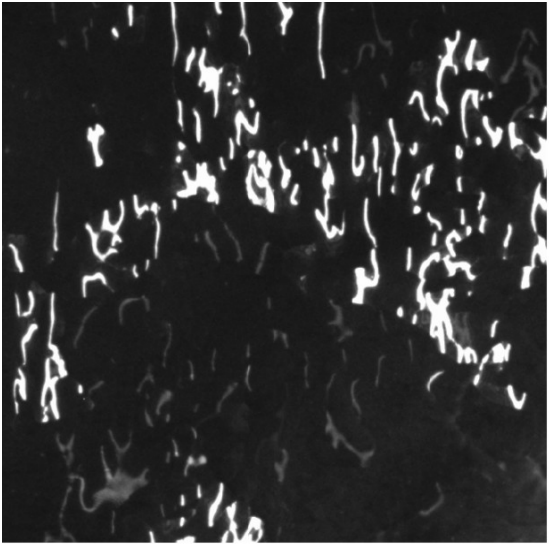

a)

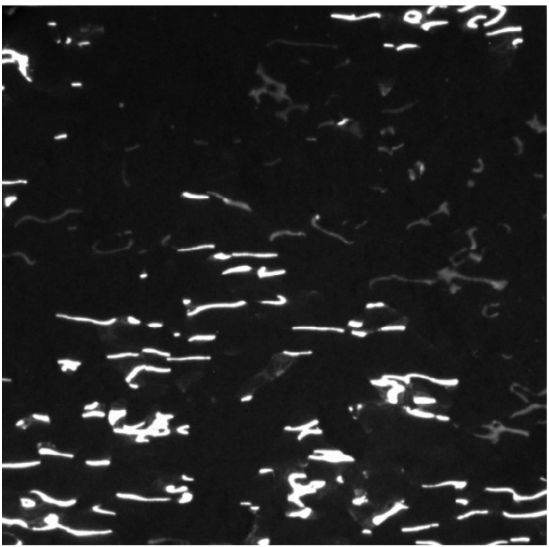

b)

**A.Alberti et al.**
**FIGURE 3 OF 5**

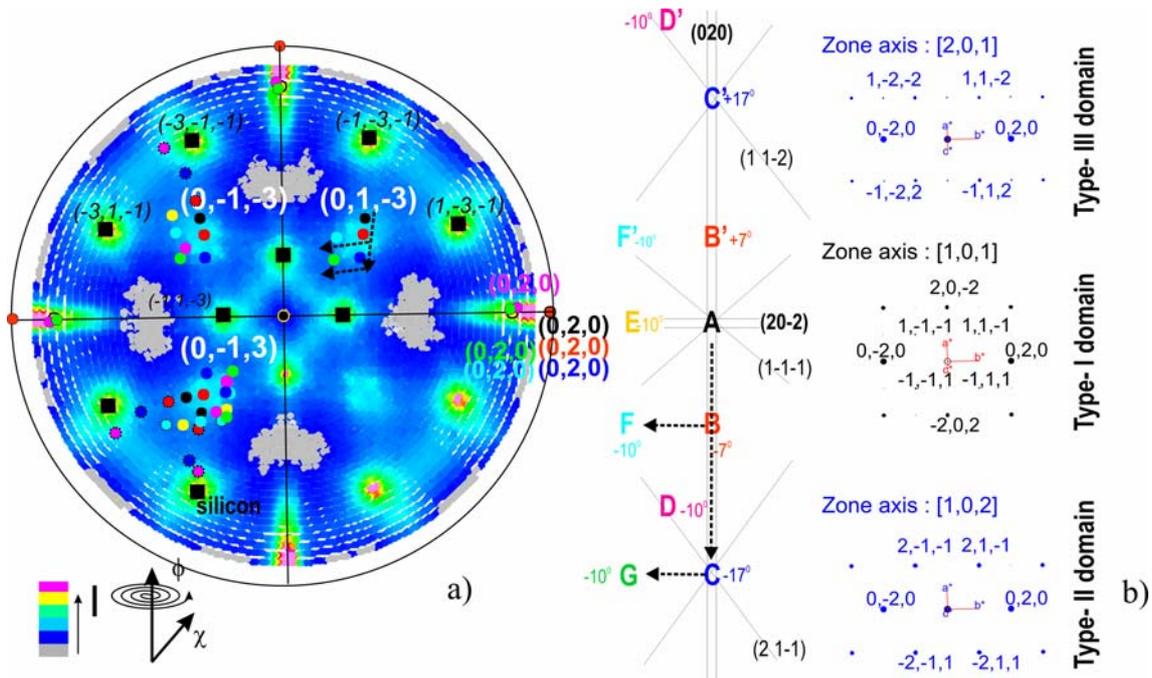

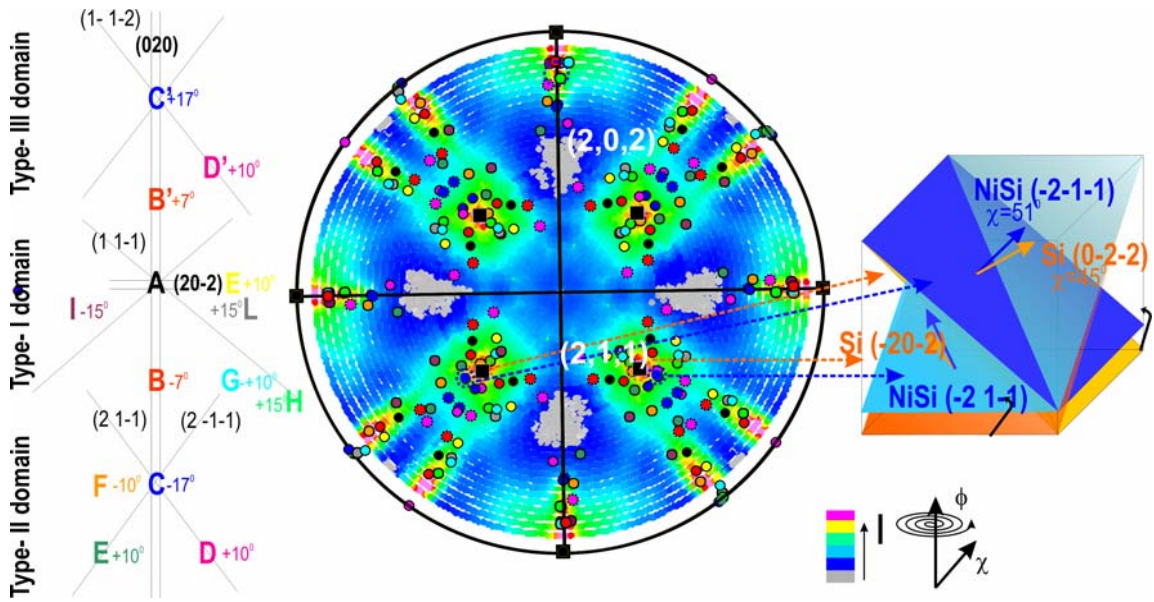

A. Alberti et al.
FIGURE 5 OF 5

**FIGURE CAPTIONS**

**Figure 1** **Structural analyses.** a) Plan-view TEM image of a 14nm-thick silicide layer reacted by spike annealing at 550°C. An intricate network of extinction contours (black lines) covers the entire area of the sample. Domains are defined by the intersection of more than two of those contours. b) Large area diffraction analysis showing the location of (020) planes of NiSi with respect to the Si lattice structure. These planes, which are perpendicular to the sample surface, have only two allowed configurations at 90° one to each other (see the dashed circles). C) and d) : plan-view image and large area diffraction analysis of a 24nm-thick NiSi layer obtained by spike annealing at 550°C..

**Figure 2** **Domain structure.** Details of type–I (a) and type–II (b) domains defined on the basis of the corresponding SAD patterns. The extinction contours are almost perpendicular to the diffraction spots labelled in the corresponding SAD, and in these regions the NiSi planes are in Bragg conditions. This conditions is lost outside the contours due to the continuous bending of the NiSi planes as sketched in (c). The bending occurs in all the directions, as represented in the case of (020) and (202) NiSi planes in (d), and it changes the properties of the interface without producing structural roughnening of the layer.

**Figure 3** **(020) NiSi bending contours.** Dark-field TEM analyses obtained by selecting each of the outlined [020] spots in figure 1b. The upper part of the sample has domains with similarly oriented (020) bending contours (a). The domains in the lower part have the (020) bending contours rotated by 90° (b). Therefore, two regions are defined on the basis of the orientation of these contours, and they cover the entire area of the sample.

**Figure 4** (a) (020)/(013) NiSi pole figure. Contributions from (131) planes of Si are also present (black squares). The lozenge-shaped patterns at $\chi=45°$ are due to (013) NiSi planes and the elongated patterns at the edges of the figure are due to the (020) planes of NiSi. (b) Schematic view of the domains forming the NiSi layer, with the corresponding diffraction patterns to identify the planes orientation. In the core of the type-I domain ("A"), the (020) planes contribute to the pole figure at χ=90° φ=0°, while the diffracted spots from (013) planes lie within the lozenges of the I and II-quarter (black circles). Moving from "A" to another regions of the domains, the diffracted spots distribute as shown by the corresponding coloured circles. Note that they accumulate in the region of the lozenges and at the edges of the figure according to the symmetry of the substrate.

**Figure 5** (a) (202)/(211) NiSi pole figure. Contributions of Si (220) planes are also present (black squares). Following the domain structure (left side schematic), some representative contributions from the NiSi lattice have been represented in the pole figure as coloured circles. Note that the (202) planes of NiSi tend to cover the features at the edge of the figure and the contribution from (211) planes accumulated within the lozenge in proximity of the Si signals and also along the splitting arms. Due to the mutual position of these planes within the lattice of NiSi, the relationship with Si is such that when one of these NiSi planes approaches the core of the lozenge, the other in the near quarter moves quite far away. This extreme configuration is not convenient. Instead, one of the most convenient location of NiSi planes with respect to the Si lattice is sketched on the right side of the figure.